\documentclass[letterpaper,11pt]{article}

\usepackage{amsmath,amssymb}
\usepackage{eepic,epic}
\usepackage{epsfig}
\usepackage{graphicx}
\usepackage{color}
\usepackage{comment}
\usepackage{tabularx}
\usepackage{booktabs} 
\usepackage{pifont}
\begin{document}

\title{The AI Liability Puzzle and A Fund-Based Work-Around}
\author{
Olivia J. Erd\'{e}lyi\\
School of Law\\
University of Canterbury\\
Christchurch 8140, New Zealand\\
olivia.erdelyi@canterbury.ac.nz
\and
G\'{a}bor Erd\'{e}lyi\\
School of Mathematics and Statistics\\
University of Canterbury\\
Christchurch 8140, New Zealand\\
gabor.erdelyi@canterbury.ac.nz
}
\date \today

\maketitle

\begin{abstract}
Certainty around the regulatory environment is crucial to enable responsible AI innovation and foster the social acceptance of these powerful new technologies. One notable source of uncertainty is, however, that the existing legal liability system is inapt to assign responsibility where a potentially harmful conduct and/or the harm itself are unforeseeable, yet some instantiations of AI and/or the harms they may trigger are not foreseeable in the legal sense. The unpredictability of how courts would handle such cases makes the risks involved in the investment and use of AI incalculable, creating an environment that is not conducive to innovation and may deprive society of some benefits AI could provide. To tackle this problem, we propose to draw insights from financial regulatory best-practices and establish a system of AI guarantee schemes. We envisage the system to form part of the broader market-structuring regulatory frameworks, with the primary function to provide a readily available, clear, and transparent funding mechanism to compensate claims that are either extremely hard or impossible to realize via conventional litigation. We propose it to be at least partially industry-funded, with funding arrangements depending on whether it would pursue other potential policy goals. We aim to engage in a high-level, comparative conceptual debate around the suitability of the foreseeability concept to limit legal liability rather than confronting the intricacies of the case law of specific jurisdictions. Recognizing the importance of the latter task, we leave this to further research in support of the legal system's incremental adaptation to the novel challenges of present and future AI technologies.
\end{abstract}

\section{Introduction}\label{Intro}

With proliferating AI-human interactions, issues around the civil and criminal liability of AI have moved to the forefront of legal policy debates. Can a bank using an AI-enabled lending decision-making system that unexpectedly turns out to unlawfully discriminate customers successfully sue the provider of the system? 
Who is liable if an autonomous vehicle (AV) hits a pedestrian or is involved in a crash? 
What happens if an AI engages in criminal actions owing to, say, an unexpected value alignment problem of the sort described in Schreier's \emph{Robot and Frank} or the canonical \emph{paperclip maximizer} doomsday scenario? 

While each of these questions touches upon different domains of legal liability---contrac\-tual, tort, and criminal liability, respectively---their core inquiry is the same: Who should be held accountable if something goes wrong with an AI and based on what rules? Well aware that courts and policymakers will soon have to come up with satisfactory answers, a growing number of papers has taken a first crack at examining the topic from various perspectives. The result is a landscape of conflicting accounts on how best to go about AI liability and the legal system's overall ability to adapt to this latest wave of technological innovation.

Some further the debate by synthesizing the relevant literature, see
Kingston~\cite{JK2016} on selected aspects of civil and criminal liability. Among those having faith in the existing system's adequacy to deal with AI liability issues is Hubbard~\cite{FH2016}, who---tacitly invoking the famous Hand Formula~\cite{Hand1947}---concludes that the current US system of contractual and tort liability strikes a fair and efficient balance between ensuring safety and incentivizing innovation. Consequently, he sees no reason to apply different metrics to compensating physical injury inflicted by \emph{sophisticated robots} (defined as having some degree of connectivity, autonomy, and potentially machine-learning (ML) ability). 

Given the apparent imminence of the topic, quite a few papers revolve around AVs. Liechtung~\cite{JL2018} urges for timely adjustment of regulation and oversight mechanisms to prepare for the impending mass-release of AVs. He also stresses the importance of clarity and predictability of the legal liability regime---whatever liability rules are chosen---so those involved in the development, production, and distribution of AVs can better assess their risk exposure. 

Several commentators argue for subjecting AVs or AI more broadly to strict liability---commonly some type of products liability regime~\cite{MG1993,JL2018}.
An interesting recent idea in this realm has been put forward by
Vladeck~\cite{DV2014}, advocating a strict liability regime entirely detached from notions of fault---in essence a court-implemented insurance system.  In a practical, goal-oriented, if slightly doctrinally inconsistent approach, he proposes to simply infer liability from negative outcomes to overcome situations where it is impossible to establish fault. Doing so, he hopes to create a more cost-efficient, equitable, and predictable liability regime, which provides a safe and stimulating environment for innovation, better protection to blameless parties, and fairer cost-spreading among affected parties. As a way of achieving the latter, he contemplates abandoning the current practice of treating AV liability as an agency question and conferring legal personality on AVs coupled with a compulsory self-insurance instead. Relatedly, Karnow~\cite{CK1994} advocates an \emph{electronic personality} for autonomous robots (by which he means those with an ML component) to enable the legal system to hold them directly liable under tort law.

A contrasting view reveals concerns about potentially ludicrous expenses involved with complex products liability suits, pre-trial settlements, product recalls, and punitive damages, pressing for a meticulous application of the negligence doctrine to AV incidents~\cite{NG2016}. He maintains that equal treatment of AVs and those under human guidance in this manner would also result in a higher degree of legal certainty spurring innovation---a common theme supporting all the above views---and allow for the operation of market-based incentives such as reputational concerns. 

Finally, Karnow~\cite{CK2016} expresses doubts as to whether any of the classic US tort doctrines---negligence and the various forms of strict liability---is up to allocating liability for wrongdoings of truly autonomous robots. This is because foreseeability is a central element of each of these doctrines, however, due to complex non-linear interactions between intricate robots and their equally convoluted, unpredictable environment, neither robots' actions nor the potential harms they may cause are foreseeable in the sense required by law. Regarding AVs and autonomous robots and mostly in the context of US tort law, other commentators have voiced similar concerns about potential liability gaps and the implications of a resulting overall uncertainty surrounding the legal liability of AI systems. They confirm Karnow's observation about the centrality of foreseeability in limiting legal liability but concede that emergent behavior (i.e., behavior contingent on the interaction of a system's elements rather than the elements themselves) exhibited by some systems may trigger genuinely unforeseeable categories of harm~\cite{WB2018,RC2018}. This unpredictability of foreseeability makes it even harder to evaluate the chances of success of litigation and hence exposure to liability, adding to the uncertainties that flow from the inconsistency of jurisprudence during the typically significant time lag needed for the legal system to adapt to novel technologies. The resulting problems---known all too well in law and technology literature~\cite{TP2018}---are inhibition of innovation and adoption of new technologies, in extreme cases reaching as far as shutting down entire emerging markets. 

In this paper, we restrict the focus of the above sketched AI liability debate by analyzing only the foreseeability concept's ability to serve as a means to limit and attribute legal liability. At the same time, however, we will also move this important discussion beyond US tort law and embodied AI systems or particular AI applications---indeed beyond any national analysis and law in general, for the following reasons:

AI is just one of the most recent waves of technological innovation (sometimes referred to as the fourth industrial revolution) all of which have fundamentally impacted our societies and economies. Due to its rapid pace of development, massively transformative nature, and other changes---most notably globalization---our world has undergone, AI is anticipated to affect humanity and our environment even more intensely. Recognizing this, there are major national AI strategies and international policy initiatives underway, which aim to forge an innovation-friendly, enabling regulatory environment, capturing benefits and minimizing potential risks AI may bring~\cite{G202019,OECD2019,EC2019,ACOLA2019}. All these initiatives converge on the point that successful societal adoption of AI---like any other form of technological innovation---requires trust on the part of society. Trust, in turn, hinges on at least some level of certainty about how AI will impact society and the economy: developers need to be able to assess the risks inherent in bringing a new product on the market, while consumers and other users of the technology must be assured that its use is reasonably safe. Without such trust and certainty, there can be no market for emerging technologies, and AI would not be the first one experiencing difficulties on this front~\cite{TP2018}. Certainty itself flows from a safe, transparent, and flexible regulatory environment that supports innovation, but designing one---even just a liability regime---is neither a purely legal nor an exclusively national enterprise.

From an economic perspective, the regulatory frameworks that structure our economies together with market imperfections crucially determine the extent to which society benefits from technological innovation~\cite{JS2015}. It is established wisdom that in our reality of imperfect markets, technological innovation is not necessarily Pareto-improving. On the contrary, in the absence of cleverly devised and potentially substantial redistributive measures, it can actually aggravate inequality and decrease overall welfare~\cite{KS2017}. Stiglitz~\cite{JS2015} also shows that inequality-related problems can only be effectively tackled by a holistic approach involving a complete and systematic revamp of market-structuring regulatory frameworks, which legal liability regimes are admittedly part of. 

The above cited key international AI policy documents, and---based on a review of relevant international relations literature---Erd\'{e}lyi and Goldsmith~\cite{EG2018} also underscore the necessity of international coordination and cooperation in the AI domain. The core of the arguments here is that issue areas with transnational impact---such as AI---are impossible to effectively regulate by means of isolated national measures. This is because their inevitable fragmentation and divergence invoke inefficiencies and tensions in international policymaking, negatively affecting domestic regimes and shattering both national and international actors' faith in such approaches.

These arguments combined call for a holistic, multidisciplinary, and transnational perspective, forbidding an isolated legal or nationally focused analysis of liability regimes. Hence, Section~\ref{Liability} will start with a high-level, comparative legal analysis,  explaining why foreseeability is central to attributing legal liability and highlighting a potential conceptual problem related to its suitability to constrain the legal liability of AI systems. We argue that foreseeability is common to all types of legal liability irrespective of the area of law they originate from, and may raise attribution problems in relation to the actions of any embodied or disembodied AI system, provided it uses certain types of ML models. Note that although under the current state of technology these problems arise in connection with certain (not all!) ML-based systems, in the future they are equally conceivable in relation to other types of AI systems. Therefore---as outlined in Section~\ref{Puzzle} in more detail---we understand the term \emph{AI system} as defined by the Organisation of Economic Co-operation and Development (OECD) in the OECD Principles on Artificial Intelligence~\cite{OECD2019}, as this allows for the inclusion of any present and future AI technologies that may pose similar challenges. We use the more generic term AI in effect interchangeably, referring to a diverse set of technologies that it encompasses at any given time. Our aim is to spur further legal debate in diverse jurisdictions. Examining the case law of multiple legal systems is, however, beyond the scope of the paper, not least because---as Barfield~\cite{WB2018} points out with respect to US case law dealing with robots---cases that could serve as precedents date back to prior to the advent of ML-based instantiations of AI and consider non-autonomous systems. Consequently, they are not directly relevant for the foreseeability problem we

Section~\ref{Puzzle} will recommend an approach to tackle current difficulties policymakers face in trying to define AI, and also engage in a short technical analysis of the current spectrum of ML-based AI systems to illustrate that a subset of them does not fulfill the foreseeability requirement. 

As noted above, prolonged legal uncertainty is just as toxic to innovation as an unduly restrictive policy stance. Hence, Section~\ref{Lia} will first suggest a minor amendment to the existing legal liability regime---as emerging based on our comparative legal analysis. This reflects our belief that---despite the appeal of such a quick-fix solution---the \emph{unaltered} application of existing liability rules to AI or a \emph{protectionistically motivated} recourse to strict liability with a view to establish responsibility at any cost are not the correct answers for three reasons: First, ignoring that those rules have been tailored to different circumstances and may hence be inept for AI, they contravene the delicately balanced objectives of the legal liability system. Second, they inhibit AI innovation by adopting an unduly punitive approach. Third, undue resort to strict liability merely circumvents foreseeability and fault problems in a dogmatically inconsistent manner rather than remedying them. 

Section~\ref{Fund} will then discuss our thoughts on the simultaneous creation of a system of AI guarantee schemes (AIGSs). This would be a clear and transparent framework for speedy compensation in cases where a liability suit has uncertain or no prospect of success owing to the unforeseeable nature of the damaging conduct, the (type of) damage itself, or the excessive costs and/or complexity of the procedure. Mirroring some aspects of financial system guarantee schemes, which form an integral part of the financial safety net and crucially contribute to maintaining trust in the financial system, the AIGSs could function as a second line of defense beyond the ambit, yet complementing the existing system of legal liability. Depending on the AIGSs' designated role within the broader regulatory frameworks structuring our economies and the policy goals they pursue, they should be in whole or in part funded by the AI industry.

\section* {Summary of Contributions:} 

\begin{itemize}
    \item Meticulous comparative legal analysis of the foreseeability concept's role and adequacy in constraining and attributing liability in the context of AI systems across \emph{all} legal domains.
    \item Multidisciplinary arguments to show the economic and political costs of failure to solve the identified foreseeability problem in a timely and internationally coordinated manner.
    \item We point out and illustrate with concrete examples that there exist AI systems that do not fulfill the legal foreseeability requirement.
    \item Recommendation on how to approach definitional difficulties in AI policy and regulatory debates.
    \item Proposal to amend existing legal liability regimes to account for AI's social utility.
    \item Proposal to set up a system of AI guarantee schemes based on financial regulation best practices to minimize legal uncertainty and foster safe and responsible AI innovation.
\end{itemize}

\section{Conditions for Imposing Legal Liability}\label{Liability}

Legal liability for AI systems could originate from either criminal or civil law. Civil liability can be further divided into contractual, tortious, and statutory liability according to the particular field of law from which the liability emanates. Conscious of the importance of striving for a globally consistent treatment of AI-related issues from the outset~\cite{EG2018}, we would like to once more note our intention to take a comparative legal approach either referring to genuinely transnational sources of law or highlighting common patterns in the law of several jurisdictions. In so doing, we hope to provide an analysis that resonates with the international community.

\emph{Contractual liability} is premised on
a contractual relationship between the parties. We illustrate how it is construed based on the 
\emph{ United Nations Convention on Contracts for the International Sale of Goods (CISG)}~\cite{UN2018}---the key international trade law convention governing the international sale of goods. The CISG is viewed as prima facie evidence for general contract principles reflecting widely accepted, international commercial best-practices making up the core of a broad set of legal systems~\cite{CB2009}. As such, it is not only derived from and influencing national legal systems, but also crucially guides international commercial arbitral tribunals, which will likely be heavily involved in the adjudication of AI-related commercial disputes in the coming years.  

In common law tradition, the CISG adopts a notion of strict liability for breach of contract: the party failing to perform its contractual obligations is liable for non-performance regardless of whether fault can be established on their part, see Articles 45 and 65 CISG and~\cite{CB2009}. However, recognizing that the party in breach cannot control all circumstances leading to non-performance, this unbounded liability is restricted in two ways. First, only \emph{foreseeable} damages can be claimed, i.e., loss the non-performing party has or 'ought to have' foreseen as a possible consequence of the breach based on information they did or 'ought to have' known when concluding the contract (Article 74 CISG). Second,  liability is excluded if the \emph{force majeure excuse} (Article 79 CISG) comes into play. Under the force majeure test---where fault becomes relevant---the non-performing party must prove that the breach was caused by an unforeseeable, unavoidable, and insurmountable impediment beyond their control. A foreseeable impediment is defined as one that parties could 'reasonably be expected to have taken [...] into account.' The foreseeability requirement determines the contractual risk allocation through an implicit, but rebuttable assumption that the party in breach assumes risk for the occurrence of foreseeable circumstances. For a good overview, see~\cite{CB2009}. Thus, in international contract law, the concept of foreseeability determines both the scope of damage claims and the extent to which liability for breach of contract can be established.

Conversely to contractual liability, \emph{tortious liability} can be triggered irrespective of whether the parties have a preexisting relationship. In fact, in line with the \emph{principle of corrective justice}, tort law links wrongdoer and victim---likely total strangers---through the notion of liability to compensate for the harm the former wrongfully inflicted upon the latter~\cite{KO2015,EW2012}. A comparison of different legal systems yields a somewhat confusing picture as to how liability is construed, but at the end of the day, countries reach similar solutions to similar problems. Our attempt to outline a systematic overview is based on~\cite{HK2012,KA2015}.

Starting with the common traits, in most jurisdictions, tort law distinguishes between negligence and strict liability, although the extent to which the latter is recognized varies considerably. \emph{Negligence} is a fault-based liability imposed on a tortfeasor that fails to exercise reasonable care, while \emph{strict liability} is negligence's no-fault counterpart, which is typically linked to the existence of a particular source of danger rather than the \emph{conduct} (either action or omission) creating it~\cite{CK2016}. Pursuant to the \emph{principle of bilateral justification} characterizing private law, a causal relationship between the tortfeasor's conduct and/or the thereby created risks and the victim's harm is universally seen as a minimum condition to shift damage to the tortfeasor and establish their legal obligation for compensation~\cite{HK2012}. \emph{Causation} is given if the harm would not have occurred but for the conduct or risks in question---this is known as the \emph{but for test} or \emph{conditio-sine-qua-non formula} in the legal jargon. However, as discussed below, all legal systems deem such an unrestricted responsibility for all damage that may ensue as a consequence of some conduct unreasonable and employ additional value judgments to confine the scope of liability. This is where the similarities stop and the inconsistencies start. 

First, there is a disturbing amount of conceptual inconsistency across various legal systems with regard to the---often interchangeably used---terms wrongfulness, fault, culpability, and negligence. As far as wrongfulness and fault are distinguished, \emph{wrongfulness}---a term widely used in Germanic countries---is an objective concept that refers to misconduct, i.e., a conduct that's somehow
incorrect in the eyes of the law, whereas \emph{fault} is a subjective notion which serves to assign blame to a certain misconduct. Systems where such a distinction does not exist follow essentially the same logic by uniting objective and subjective considerations under the umbrella of a single term (fault in France) or resorting to additional helping mechanisms (e.g., duties of care in English and US law). English law views fault as an element of wrongfulness, yet tends to use both concepts interchangeably. US law makes no distinction at all and prefers the term \emph{culpability} or fault over wrongfulness. \emph{Negligence}---popular especially in common law jurisdictions---either means the fault-based class of tort liability by contrast to strict liability or is used as a synonym for fault and culpability. 

To make sense of this, it is helpful to acknowledge that all legal systems aim to protect various rights and interests by identifying and preventing potentially harmful and hence wrongful behaviors. There is, however, a key difference in how civil and common law jurisdictions go about this: Following a more systematic approach, civil law countries have chosen to codify such behaviors---which form the basis of wrongfulness---in distinct statutory provisions. In common law systems, on the other hand, such standards of conduct have been incrementally developed through case law by defining specific duties of care for different types of torts. Correspondingly, to hold a defendant liable for damages caused by their conduct, civil law's wrongfulness or fault inquiry focuses on whether the factual elements of a norm have been fulfilled and---if the norm in question establishes fault-based liability---whether the conduct should be qualified as careless under the given circumstances. Common law approaches torts in a slightly different, yet in terms of the outcome essentially 
similar manner: Responsibility for strict liability torts merely requires proof that a particular harm occurred, that said harm was caused by the defendant's conduct, and that the defendant could foresee at least the type of harm that transpired, while in case of negligence torts an additional breach of a particular duty of care by a faulty or negligent conduct is necessary.

Jurisdictions also differ in how they measure fault and negligence. The prevalent objective standard of measurement considers a conduct faulty or negligent if it lacks \emph{reasonable} or \emph{ordinary care}, i.e., does not correspond to the way a reasonably prudent person would have acted in the defendant's position. Most strikingly in the US, the negligence standard is an economically charged concept determined by a balancing approach---essentially an economic cost-benefit analysis---known as the already mentioned Hand Formula~\cite{Hand1947}: A conduct is deemed negligent if the \emph{expected} harm---the magnitude of a potential loss (L) adjusted by the probability of its occurrence (P)---outweighs the costs to avoid the harm---the burden of undertaking precautionary measures (B). Put formally, a duty of care is generated where $P \times L > B $~\cite{RP1972,BW1990}. 
Other common law jurisdictions rely on this economic logic more covertly and often include additional factors, like the social utility of the conduct, among the balancing criteria, in effect modifying the above formula as follows $P \times L > B + U$, where U stands for social utility. This objective approach is usually justified with reference to the exorbitantly high administrative costs of determining each defendant's abilities on an individual basis, the observation that the tortfeasor's abilities have no bearing on how their actions affect others, the endeavor to reinforce people's moral responsibility, or that the law needs to define average standards of conduct in pursuit of general welfare. Proponents of a subjective approach criticize that this amounts to an imposition of strict liability in cases where the defendant's abilities are below average. There is little practical difference between the two approaches, as ultimately both require courts' discretionary judgment on whether it is reasonable to impose liability on a case-by-case basis. 

As pointed out earlier, all jurisdictions reduce the scope of liability delineated solely through causation. Here again, approaches and terminology are confusingly inconsistent both across jurisdictions and different points in time, but restrictions are achieved in two basic ways: By limiting either  
\emph{causation} or 
the \emph{scope of liability}. The first technique works with an unbounded notion of fault imputing liability for \emph{all} damages caused by a conduct, but treats causation as a normative rather than natural concept and employs the \emph{theory of adequacy} to exclude liability for \emph{atypical} or \emph{remote} damage, i.e., damage stemming from an entirely coincidental, objectively unforeseeable interplay 
of circumstances, which the tortfeasor could not have possibly controlled. The second method conceives of causality in the natural sense of the term and---based on prediction theory---limits the scope of liability by restricting the duty of care to \emph{foreseeable harms}, i.e., those the defendant should have actually been capable to avoid~\cite{CK2016}. It follows that, either way, fault based liability can only be imputed for foreseeable harms. 

Perhaps less intuitively, foreseeability is equally central to strict liability torts, despite the fact that fault plays no role here. To understand why, consider that strict liability is imposed on the premise that someone creates a source of danger, which is likely to cause harm and---crucially---which said person has the \emph{ability to control}. Yet control implies that both dangerousness and potential harms are recognizable, that is, foreseeable. US doctrines of strict liability include ultrahazardous activity and three types of products liability. In civil law systems, a number of specific statutory provisions prescribe non-fault based liability for keepers of, e.g., animals and motor vehicles, and other hazardous activities. Jurisdictions with a strict-liability-averse stance, such as England, solve such cases over negligence, but they tend to stretch duty of care requirements so far that liability becomes virtually inevitable---yet another example of similar results achieved by seemingly distinct approaches. 

Similar arguments support the claim that foreseeability is also an essential condition for the imposition of statutory liability: Statutes pre- or proscribe a certain conduct to prevent some risks typically inherent in that behavior from materializing, whereas the scope of a norm cannot reach beyond the limits of foreseeability. Instruments like the \emph{protective purpose theory} in some European legal systems or the \emph{harm-within-the-risk rule} in the US serve the purpose to limit liability for breach of statutory provisions based on this logic.

Turning to criminal liability, we now outline the basic requirements for establishing criminal responsibility based on the comparative analyses of Anglo-American, continental, and international criminal law provided by Fletcher and Marchuk~\cite{IM2014,GF2000}. Pursuant to the \emph{legality principle}---a central moral principle of criminal law expressed by the Latin term \emph{nullum crimen sine lege (no crime without law)}---criminal punishment typically presupposes that a particular conduct is criminalized by law, i.e., penalized behavior and potential sanctions---the severity of which reaches well beyond those imposed under civil law---are clearly laid down in statutory provisions. Criminal law pursues primarily punitive objectives against those engaging in statutorily criminalized socially unacceptable behavior while being mentally capable to recognize the unlawfulness of their conduct. Committing a crime always requires a physical element referred to as \emph{actus reus (guilty act)}. With the exception of strict liability offenses, where the blameworthiness of a conduct that violates a norm protecting certain societal values is presumed, this must be accompanied by a subjective element referred to as \emph{mens rea} (a.k.a. \emph{culpability}, \emph{fault}, or \emph{blameworthiness})---criminal law is also plagued by a fair amount of terminological inconsistency. By contrast to tort liability, which takes recourse to mostly objective standards to determine the blameworthiness of a conduct, criminal law measures the defendant's mental state by a predominantly subjective test.

Mens rea encompasses a range of different mental states 
described by a bewildering variety of terms both within and across legal systems. The three broad categories distinguished are intent (\emph{dolus}), recklessness, and negligence (\emph{culpa}). \emph{Intent} is commonly divided into two---in certain legal systems three---subcategories: (1) \emph{Dolus directus of the first degree} (\emph{direct} or \emph{specific intent}) requires \emph{purposeful} conduct, i.e., that a person commits an offense with the desire to achieve a particular prohibited result. (2) \emph{Dolus directus of the second degree} (\emph{oblique} or \emph{general intent}) is given if an offender acts \emph{knowingly}, that is, intends to commit a 
prohibited act without desiring to achieve a specific harm but foreseeing its occurrence as virtually certain. The differentiation between these two forms of intent is not always present: for instance Article 30 of the Rome Statute of the International Criminal Court~\cite{RS1998}---a central part of the body of international criminal law---requires the cumulative presence of both volitional and cognitive components. (3) In addition, especially continental criminal legal systems stipulate a third, more indirect notion of intent called \emph{dolus eventualis}, which focuses on the offender's attitude towards the consequences of their action and is satisfied if an individual remains \emph{indifferent} despite foreseeing a possible harm. 

\emph{Recklessness} is an intermediate form of culpable state between intent and negligence in common law jurisdictions, which penalizes behavior that grossly deviates from the standard of conduct of a reasonable person. It is given if an offender is aware of, yet \emph{consciously disregards the substantial and unjustifiable risk} that their conduct will have negative consequences. It is a volitional element without an equivalent in civil law systems, although conscious negligence (explained below) can be regarded as its closest counterpart.  

Like in tort law, \emph{negligence} in criminal law also connotes a behavior that departs from the objective standard of conduct of a prudent person. Ordinarily, 
negligence lacks a cognitive element---which is why English and US law are divided on whether it
counts as a class of mens rea---i.e., the offender \emph{should have been, but was not aware of the substantial and unjustifiable risk} that their action may have negative consequences. Beside this \emph{unconscious negligence}, some jurisdictions distinguish a second form of negligence dubbed \emph{conscious negligence}, given if a person \emph{foresees the risk of 
harm} \emph{but believes}---indeed almost hopes---\emph{it will not occur}. To justify the imposition of significantly weightier sanctions, criminal law usually requires \emph{gross negligence}, i.e., considerable deviation from the reasonable person standard. It is fulfillled if an individual's actions pose an \emph{obvious risk} to bring about \emph{substantial harm} and the offender has the \emph{ability to take precautionary measures}. Moreover, negligence is typically only penalized if explicitly criminalized by law---a case in point being Article 30 of the Rome Statute, which excludes criminal responsibility for negligent behavior unless other provisions of the Statute expressly so provide. 

Hence, with the exception of strict liability and unconscious negligence offenses, criminal responsibility likewise presupposes that the offender foresees the potential harms their conduct may cause. In those two particular cases, however, only behaviors explicitly criminalized by law entail liability, and such statutory provisions are only conceivable if the legislator foresees that the conduct may result in harm. 

In conclusion, we can observe that foreseeability (reflecting an inherent ability to control) features prominently among the conditions for imposing any type of legal liability. Admittedly, case law in disparate jurisdictions and legal domains adds a number of convoluted facets to this problem, but for now we would refrain to get into those issues. The important insight at this initial stage is to realize that we face a general legal problem, which spans jurisdictions and legal domains, has potentially severe economic and political implications, and consequently needs to be addressed as soon and as widely as possible. On this note, let us now investigate if and to what extent AI is foreseeable and controllable in the sense required by law.    

\section{Foreseeability: the Missing Piece of the AI Liability Puzzle}\label{Puzzle}

 Academic papers, policy documents, and other contributions discussing various aspects of the regulatory treatment of AI typically either handle the concept as given and thus refrain from defining what they mean by AI, or choose a working definition that is best suited to their particular inquiry. While this is not surprising---after all, to date a universally accepted AI definition does not exist---it does create problems of definitional inconsistency. This, in turn, curtails efforts to clearly distinguish between various technologies referred to under the common banner of AI, identify their essence and most relevant properties for distinct policy purposes, and establish an internationally consistent policy stance towards them. 
 
 At present, much of the energy dedicated to defining AI is directed to finding some one-size-fits-all definition that is universally applicable in any given context. Insofar as this fosters consistency, we applaud this intent. However, looking at things through a teleological lens highlights that any such broad definition is only of limited use. The purpose of defining AI one way or another is to create a concept with clearly identifiable attributes that we can understand, allowing us to assess AI's capabilities, anticipate its actions and the consequences of those actions, and---ultimately---to make informed decisions on what roles we want it to play in our societies. Yet AI is an umbrella term, which may refer to a number of very different notions (from AI as a scientific field~\cite{PM2017}, to sub-fields of AI like robotics and ML, to specific technologies like differing ML models), the range of which varies over time (recall, for instance, that what we now know as \emph{big data} was considered AI a few years back). It is apparent that these concepts and/or technologies exhibit distinct characteristics and serve very different purposes, so that they cannot be treated as a single, homogeneous thing. 
 
 We are of the opinion that across the manifold use cases and contexts in which we may encounter AI technologies, the technologies themselves (as determined by their state of art at a given point in time) are the only constant elements. The capabilities of an AI and consequently the tasks it may take on in human societies (if we so decide) are also determined by the technology it uses. We therefore believe that any AI definition developed for regulatory purposes should first and foremost be based on the particular technology in question. That said, in the presence of such specific definitions, it may make sense to additionally apply a few more generic definitions, to provide for consistency in selected domains.

 In this paper we chose to use the term AI system---which we believe is a good example for a generic definition that promotes consistency---as defined in the OECD AI Principles to delineate the group of AI technologies that form the subject of our inquiry. The OECD defines \emph{AI system} as a 'machine-based system that' may operate at varying levels of autonomy and 'can, for a given set of human-defined objectives, make predictions, recommendations, or decisions influencing real or virtual environments'~\cite{OECD2019}. Here we only consider ML models, a particular class of AI systems. Expanding the taxonomy provided by Flach~\cite{PF2012}, we distinguish four main ML models, namely geometric, probabilistic, logic-based, and neural networks (NNs). These four types of model classes form a sort of continuum with logic-based models---which typically have no foreseeability issues---on one end and NNs---which pose the biggest obstacles regarding foreseeability---on the other. Since our goal is merely to point out that \emph{there exist} AI systems that present foreseeability issues (rather than to give an exhaustive list of AI technologies that do so) we will only concentrate on NNs to provide such an example.

As a preliminary matter, both policymakers and society at large need to be conscious of the fact that AI does not \emph{know}, \emph{think}, \emph{foresee}, \emph{care}, or \emph{behave} in the anthropomorphic sense, rather it applies what could be best described as \emph{machine logic}. To illustrate the potential implications of that distinction, consider the following example: 
 ML-based systems---which raise the biggest technical and legal challenges due to their unpredictability stemming from their independent learning property---do not \emph{know} why a given input should be associated with a specific label (e.g., that a small, red, circular object is a ball), only that certain inputs are \emph{correlated} with that label~\cite{ZL2018}. That is, the system identifies outputs based on a set of predefined parameters and probability thresholds through a process that is fundamentally different from human thinking. 

Conventional ML-based systems usually use \emph{human engineered} feature extractors to process raw data in order to receive a suitable representation the system can work with. By contrast, deep learning (DL)-based systems---a neural network-based subgroup of ML approaches---are capable of processing raw data on their own, automatically identifying the right representation they need for classification. Furthermore, DL-based systems do not just use this one representation, but possess a nested hierarchy of representations obtained by transforming a lower level representation (starting with the raw data) to a higher, more abstract level of representation~\cite{Lec:j:DL-nature}. 

What is more, this type of machine reasoning always implies a certain probability of failure, where failures tend to occur in---from a human perspective---unexpected ways and may have different reasons. Let us give three examples. 

In the first example, the failure is caused by a \emph{bad classifier} as illustrated by Ribeiro et al.~\cite{RSG2016} in their Husky vs. Wolf experiment. Here, a system trained with 10 wolf and 10 husky pictures was given the task to distinguish between wolves and huskies. On purpose, all wolf pictures had snow in the background but none of the husky pictures. Since snow was a common element in the wolf pictures but was not present in the husky pictures, the system regarded snow (or better, white patterns on the lower parts of the pictures) as a classifier for wolves. Thus, in the experiment the system predicted huskies in pictures with snow as wolves and vice-versa. The second example shows a way of cheating a facial recognition system (FRS) by a \emph{physically executed attack} (i.e., manipulating the physical state the system analyzes rather than its digitized representation) introduced by Sharif et al.~\cite{SBBR2016}. FRS's are usually using neural networks 
in order to recognize patterns in big datasets, in this case the differences between millions of faces (e.g., the relative position of nose and eyebrows, size of the nose, etc.). They used a pair of glasses with a colorful frame which basically interfered with the system's pattern recognition. It not just blocked the \emph{view} to crucial parts of the faces but, due to the colorful frame, gave the system the impression that it sees some patterns and so the FRS often made mistakes despite indicating a high probability of confidence. Our last example shows an \emph{adversarial setting} (essentially an automated attack on a search algorithm to minimize its utility) where the adversary's goal was to create inputs that a DL-based system misclassifies, however, humans do not~\cite{sze:tr:intriguing}. Their adversary system manipulated input data by adding what is called \emph{noise} not detectable to human eyes to the original pictures, fooling a DL-based system into classifying a school bus and a pyramid as an ostrich.

Even without going 
into technical details, the above analysis shows that it is conceivable that AI systems and the way they generate failures are too complex or otherwise impossible to anticipate and hence do not satisfy the legal foreseeabilty requirement. This insight suggests that we have to develop new legal solutions to attribute liability to such AI systems---or more precisely, given they lack legal personality, their designers and/or manufacturers---as well as to distribute liability between them and their human operator(s). To tackle this challenge, regulators and policymakers will need to dynamically determine which particular AI technologies pose forseeability problems at any given juncture---a task that, in our view, will involve developing and continuously adapting specific definitions tailored to each of those technologies.

\section{Adapting the Legal Liability Regime}\label{Lia}

As shown in the previous sections, holding AI systems---again, until such time that they acquire legal personality, only the people contributing to their design and/or distribution---liable for infringement of legal rights and the resulting damages may not be possible because the failure causing the harm and/or the harm itself were not foreseeable. 
This leads to considerable uncertainty, which, in turn, significantly hinders AI innovation, as well as trust in and social acceptance of AI technologies. It is a serious problem that needs urgent solution, unless we want to delay or even miss out on the economic and social benefits AI could bring to humanity. Note that the extent and distribution of aggregate benefits are conditional upon handling AI innovation the right way, especially in a welfare-enhancing rather than economic-inequality-aggravating manner~\cite{KS2017} and mindful of dual-use concerns~\cite{Malice2018}. 
 
Yet, any reform proposal should carefully balance the objectives a particular field of law seeks to pursue by the imposition of liability and the overarching policy objectives guiding AI innovation. 
Contract law, where liability for non-performance ultimately aims to achieve an optimal allocation of contractual risks, is the least problematic area, as leaving the matter of contractual risk allocation---through, e.g., negotiation of appropriate guarantee arrangements---to the parties' free disposition will usually yield fair results even with AI's unpredictability. For instance, a seller's willingness to assume risk for an unforeseeable failure of an AI system sold to a buyer can be offset by a higher price negotiated. That said, the framing of policy debates on AI may influence parties' expectations and there may be scope for regulation to correct unjustified fears or overreactions.

As regards tort law, compensation and deterrence are widely recognized as its main policy goals. Loss-spreading, vindication of rights, denunciation of wrongdoing, and educating the public on the proper standards of conduct are mentioned as auxiliary objectives, while views differ on whether private law can or should pursue public interest or have punitive functions~\cite{HK2012,KO2015,GC2015,RP1972}. Following in the footsteps of the school of economic analysis of law, many regard civil liability as an instrument to regulate safety~\cite{RP1972}. This opinion is supported by the fact that tort law falls within the remit of a distinct regulatory strategy recognized in regulatory theory labeled \emph{allocation of rights and responsibilities}~\cite{BC2012}. In this view, negligence incentivizes safe conduct up to the economically efficient level (although admittedly without regard to available alternative activities), while strict liability both regulates the socially desired level of hazardous activities and encourages safety~\cite{RP1972,GC2015}. An often-heard criticism in the legal community is that such objective economic efficiency analyses are blind to equity considerations and hence not reconcilable with the nuanced legal analysis that is necessary to ensure fair and just outcomes. However, this view seems to ignore that the application of any such analysis requires prior policy choices on which values to maximize and which losses to minimize in order to maximize society's welfare. So while the analyses as decision-making tools are admittedly value-neutral, their use is always preceded by a set of very much value-laden and equity-driven legal and policy judgments~\cite{BW1990}. Over time, such judgments has set duties of care for particular negligence torts and designated sources of danger to be addressed by strict liability torts reflecting a careful and well-established balance of contradicting interests.

Resonating with Hubbard~\cite{FH2016} on the necessity of keeping this delicate balance, we therefore think that it is mistaken to succumb to the temptation to bypass fault or foreseeabilty problems by classifying all 
instantiations of AI as dangerous and \emph{punishing} them with the overarching imposition of strict liability. This would not only be dogmatically incorrect, but would also strangle AI innovation and reduce social welfare compared to an ideal,  hypothetical alternative. In instances where the foreseeability requirement is satisfied and hence duties of care can be determined, we propose to include in the Hand Formula a further variable (U) capturing the economic and social utility derived from AI innovation and it's progressive adoption in society such that $P \times L > B + U$. Considerations on the social utility of an activity should also be included into courts' overall balancing exercise in jurisdictions, where a less mathematically explicit approach has been established. This solution would require development of new methods to quantify, measure, and allocate such gains to AI innovators, producers, and distributors instead of leaving it up to courts' discretion. Furthermore, these gains should also be accounted for when establishing potential strict liability standards for AI. 

As for criminal liability, hardening liability standards by circumventing the foreseeability requirement is not reconcilable with justifying the imposition of criminal sanctions.

So yes, it is certainly desirable that AI engineers have complete \emph{control} over their systems and avoid design and training failures. But is this a realistic expectation at the present juncture? More importantly, are we willing to stall the adoption of AI until we can \emph{guarantee} its safety? Or is there a compromise that encourages both reasonable safety and innovation?  

\section{AI Guarantee Schemes as Work-Around}\label{Fund}

All this leaves us with the problem that, even assuming the legal system will incrementally adapt and solve the above highlighted foreseeability loopholes and other challenges posed by AI, this will take time. As noted by Pearl~\cite{TP2018} in respect of AVs and the US tort system, we are probably looking to several decades of deliberation, trial-and-error type of progress in the legal treatment of AI, and inconsistent jurisprudence. Yet legal certainty is indispensable to get the most out of AI. Again, this problem is not specific to AI but common to all new technologies, and there have been a number of other instances over the course of history, where fears about the legal system's ability to rise to certain challenges have prompted a search for alternative solutions and regulatory interventions on the part of the state. 

Examples include \emph{no-fault insurance-based solutions}, which substitute for and eliminate access to the judicial system. Such accident insurance schemes are in place in several countries in diverse fields like occupational, medical, and all types of personal injuries. Dispensing with the need to examine how the damage occurred, these systems guarantee victims fast compensation of their claims and involve lower administrative costs compared to litigation. However, they have the negative effect of promoting carelessness. Moreover, due to financial constraints, they typically only offer partial compensation through the introduction of arbitrary restrictions conversely to the judicial system, which, provided successful litigation, fully compensates victims. Measures to alleviate these weaknesses---such as making the amount of compensation conditional on the specific circumstances under which the damage occurred or granting insurers rights of recourse against tortfeasors---help to provide a fairer and more equitable compensation, but do so at the cost of speed and increased costs due to the necessary legal inquiry into causation. Critics of no-fault insurance-based systems, therefore, see little practical difference to litigation and advocate that they complement rather than substitute tort law. See Koziol~\cite{HK2012} for more details.

Another approach, analyzed by Pearl~\cite{TP2018}, is setting up \emph{victim compensation funds}---cases in point are the 9/11 Victim Compensation Fund and the Gulf Coast Claims Facility established after the Deepwater Horizon disaster---which exist parallel to rather than in lieu of the judicial system. Victim compensation funds have been implemented on multiple occasions with varying objectives such as relieving pressure on courts, supporting ailing industries, or simplifying and expediting compensation processes. As regards design questions, they may be established either as quasi-judicial or non-judicial funds. \emph{Quasi-judicial funds} are administered by the judicial system or a public agency and financed by taxes or fines imposed on a selected group of individuals or organizations, who would assume a defensive position in the event of litigation. \emph{Non-judicial funds} are divided into three sub-categories: \emph{Public funds}, which are administered and at least partially funded by government or an entity with government authority, 
\emph{private funds}, administered and funded by private organizations, 
and \emph{charitable funds}. The latter are also privately administered and funded by private donations, yet they are distinct from the other three types of funds in two respects. First, their only purpose is to minimize administrative and logistical burdens of distributing donations rather than providing an alternative to litigation. Second, they tend to provide flat compensation awards without recourse to tort law to determine eligibility for compensation. One notable advantage of victim compensation funds over conventional litigation is flexibility, given that their status, funding, administration, and processes are customarily designed with a particular set of circumstances in mind. As a general rule, they are also a faster, more efficient, and cost-effective alternative to the tort system. Against these advantages weigh the potentially massive administrative burdens of establishing funds, which by far outweigh their counterparts in the judicial system. Victim compensation funds typically also fall short of providing the degree of transparency and publicity inherent in conventional litigation---attributes that may be of core importance to victims. 

Outlining the significant benefits involved, Pearl recommends the creation of a fund for AV crash victims in the United States at least until the legal system catches up with AI innovation. She proposes the establishment of a quasi-judicial fund administered by the National Highway Transportation Safety Administration (NHTSA) and funded by taxes on the sale of AVs to be paid by sellers and purchasers. Requiring both buyers and sellers to contribute, she argues, is justified by the fact that both groups would benefit from the introduction of AVs. She envisages a voluntary participation both for victims (requiring them to file a claim with the fund and waive their right to litigation upon acceptance of the compensation award) and AV manufacturers (under the condition of paying their share of the AV sales tax and participating in data-sharing and design improvement programs). Finally, the fund should only cover human injuries and fatalities, whereby compensation should be full and automobile insurance companies (whose subrogation rights would be extinguished where victims accept compensation awards) should be allowed to seek reimbursement from victims' compensation awards to recover any prior insurance payouts.

Our proposal to create a system of \emph{AIGSs}  
is inspired by the various types of guarantee schemes (most notably deposit guarantee, insurance guarantee, and investor compensation schemes) used in the financial system (hereinafter FGSs.) FGSs are usually sectorally configured, at least partially industry-funded, and sovereign-backed guarantee funds. Together with a number of other arrangements---such as lender or market maker of last resort support from central banks---they make up the heterogeneous group of \emph{financial system guarantees}. Broadly speaking, these guarantees (sometimes also referred to as \emph{financial system safety net}) are designed to provide assurance to those involved in financial transactions with financial institutions or markets that their claims against their counterparties will be met even in the event of a major liquidity shock or failure of the latter. Heavily expanded in the wake and after the global financial crisis, they are a widely used and successful model to safeguard financial stability by preserving confidence in the financial system in times of stress~\cite{KD2004,SK2011}. 

Even though perhaps most prominent in the context of financial markets, this powerful feedback-loop between the extent of uncertainty and the level of trust is a central determinant in shaping any market, AI being no exception. So, to foster confidence, the idea is for the AIGSs to provide a transparent, predictable, and reliable alternative funding mechanism outside of the scope of the legal liability system to compensate aggrieved parties. Compensation should be available in a contractual, tort, or, as appropriate, criminal context in cases where legal liability cannot be established due to the lack of foreseeabilty of an AI performance failure and/or the resulting harm. Furthermore, the AIGS should also be available to shore up the legal system in the face of the anticipated uncertainty and complexity of AI-related litigation while courts and policymakers grapple with other novel problems arising from AI. Because such difficulties are likely to occur worldwide and in all domains impacted by AI, our proposal is geared to the global context, taking a country- and domain-neutral approach. Our goal is once again to spark a high-level, conceptual debate that can inform future 
policy initiatives in this domain. 

Governance arrangements of any guarantee scheme are strongly dependent on the broader governance structures adopted in industries to which they are linked. Given the preliminary stage of discussions on AI governance in virtually any domain and country, it is relatively hard to define robust design criteria. Nevertheless, based on Davis'~\cite{KD2004} survey of international practices with respect to FGSs and Pearl's above recommendation on a US national AV victim compensation fund, we will attempt to sketch out an initial set of principles to guide future deliberations on this issue. 

\emph{Nature of the scheme}: Beyond the obvious motivation to provide predictability regarding compensation, we see AIGSs as integral parts of the broader domestic and eventually global AI governance frameworks, which pursue the overarching objective of ensuring that the development and adoption of AI technologies are beneficial to humanity. One facet of that endeavor is to incentivize AI innovators to employ responsible and safe practices, but the funds could also be instrumental in furthering other policy objectives, such as mitigating AI's inequality-aggravating impacts by redistributing some of the costs and benefits of AI innovation. In light of these strong public policy implications, quasi-judicial funds do indeed seem best suited to function as AIGSs.   

\emph{Administration}: Among the ranks of academics and various public and private organizations and groupings vested with AI policy development, there is a growing consensus about the necessity of some sort of global governance framework for AI at some point in the future~\cite{EG2018,AKEPRS2018}. However, at least at the present juncture, the reality is that AI innovation and implementation is outpacing policymakers' regulatory and oversight capabilities. Countries are busy weaving their national AI strategies and passing the most important pieces of legislation to have at least some semblance of control over the most pressing issues across diverse policy domains---like healthcare, financial services, the criminal justice system, or welfare---without much regard to cross-sectoral consistency. As the 2018 AI Now Report~\cite{AINow2018} acknowledges, each of these distinct domains has its established regulatory frameworks, traditions, and specific difficulties, requiring specialized expertise and sector-specific regulation. This and nascent national practices suggest that AI governance will initially be structured in a domain-specific fashion with existing agencies taking on AI-related regulatory functions. Given the need for speedy policy response, this is a commendable approach at least in the interim, until more research can be done on the optimality of governance arrangements. In the context of FSG governance, the Davis report~\cite{KD2004} identifies six key governance objectives, stressing that governance arrangements should (1) establish clear lines of responsibility avoiding duplication of regulatory mandates, (2) eliminate avenues for conflicts of interests, (3) keep the administrative costs of the fund as well as (4) compliance burdens for industry as low as possible, (5) where appropriate, involve industry stakeholders, harnessing their expertise, and (6) provide an adequate incentive structure for regulatory authorities. 

Taken together, these observations furnish strong arguments to house AIGSs within domain-specific agencies, at least until experience provides us with more clarity on the vices and virtues of such an approach. In the meantime, we strongly encourage the international community to keep up efforts towards setting up a global AI governance framework---preferably involving some element of self-regulation to benefit from multifaceted expertise and ensure a truly dynamic and recursive whole-of-society dialogue. Once up and running, such cross-jurisdictional governance arrangements could justify a transnationally organized AIGS system. Although a glance at financial regulatory experience gives reason to doubt the practical feasibility of any sort of global plans and we realize that the prospect is in any case a remote one, it should still not be entirely taken off the table.

\emph{Coverage}: As noted by the Davis report~\cite{KD2004}, fund coverage design inevitably involves wrestling and eventually putting up with tradeoffs between the conflicting objectives of efficiency, equity, and minimum complexity and cost. Note that the costs of guarantee schemes are not restricted to the amount of compensation paid out, but also include potentially significant administrative and compliance costs (e.g., of the establishment, ongoing operation of schemes, and dispute resolution mechanisms) and much less obvious indirect costs to society in the form of moral hazard and related behavioral problems. The appropriate balance between different objectives is typically sector-dependent and tools like coverage limits, coinsurance, and means testing are among those employed to find a suitable configuration. In widespread opinion, in view of guarantee schemes' role as safety net (a sort of back-up solution), they should ideally only step in to compensate substantial losses. Given the abundance of unknown variables in this respect, we would refrain from offering any specific recommendation at this time.

\emph{Participation}: In theory, participation in guarantee schemes may be either voluntary or compulsory. Nevertheless, few FGSs leave this matter to financial institutions' discretion. Instead, they typically foresee compulsory participation to avoid problems of adverse selection (i.e., disproportionate representation of the least reliable institutions in funds). This argument also holds for AI innovators' and manufacturers' recourse to AIGSs, suggesting that compulsory participation may in fact be preferable. Such an approach could additionally be justified by AIGSs intended rational as a tool to regulate the AI industry's incentive structure, while at the same time potentially pursuing other policy objectives. 

\emph{Funding and pricing}: Guarantee schemes involve a redistribution of losses, calling certain stakeholders to foot the bill to alleviate pressure on others. Striking a level of redistribution that stakeholders perceive as fair is therefore key to ensure guarantee funds' acceptance and efficiency. Funding relates to the timing and rate of contributions, as well as the base of funding, while pricing determines contributors' relative share. With respect to FGSs, the Davis report notes that funding and pricing considerations should strive to accommodate four general goals: (1) cost efficiency (minimize administrative costs), (2) competitive neutrality (equitable treatment of contributors with similar characteristics), (3) stability (predictable and broadest possible funding base), and (4) allocative efficiency (eliminate moral hazard incentives). 

In terms of the timing of funding, fund administrators have the choice between pre-funding (where contributions are paid into and managed by the fund), post-funding (whereby contributors incur contingent liabilities and are only required to pay into the fund after the guarantee triggering event), or a combination of both. Pre-funding usually implies greater stability and credibility that funds are readily available in the event of a crisis. It is also conducive to a higher acceptance of risk-sensitive pricing, typically perceived as fair, and requires less financial back-up by the public purse. On the down side, pre-funding may lead to higher than warranted contributions due to the uncertainty of triggering events' occurrence. It may also create moral hazard incentives, raise issues around controlling the size of the contribution pool, and be less cost efficient than post-funding. Post funding, on the other hand has a pro-cyclical impact, in that it imposes a burden on contributors after a guarantee event, compounding their financial difficulties. 

Regarding the funding base, the main questions revolve around the relative ratio of public and private funding, whether to establish several domain-specific schemes or one cross-sectoral fund, and the basis for calculating contributions. The pros of domain-specific schemes include cost efficiency, competitive neutrality, sensitivity to domain specific characteristics, and avoidance of cross-subsidies. However, they are less financially stable, have a restricted ability to realize diversification benefits, and may face transition problems due to structural changes in the organization of contributing entities. 

Finally, pricing choices are usually about striking an acceptable balance between simplicity and efficiency. The latter is promoted by differential, risk-sensitive contributions, which are typically the better choice when it comes to combating moral hazard and ensuring equitable treatment of contributors, but are also complex to implement. The alternative is to require uniform, flat-rate contributions, which excel in simplicity, transparency, and involve low implementation costs. 

Applying these insights to AIGSs, the kinds of systemic crises with the potential to deplete FGSs and necessitate state involvement are admittedly a highly remote possibility in the AI context. However, since it is impossible to predict the exact trajectory of AI innovation, it is hard to anticipate if and how this might change in the future. This uncertainty coupled with better feasibility of risk-sensitive pricing and the likelihood that industry would perceive pre-funding as the fairer funding option are strong arguments in favor of pre-funding. Because of the lack of large-scale shocks that may strain schemes' funding resources, it is unlikely that post-funding, even in an auxiliary form, will be necessary (again, this may change based on how the current state of affairs develop). As for funding base, we believe that, unless specific policy considerations dictate otherwise, this should be restricted to private contributions from the AI industry---the group whose incentive structure it aims to target---without involving public funds or contributions from AI users. Recalling our above recommendation for domain-specific AI governance arrangements, funding should be organized on a sectoral basis. Contributions should be calculated taking due account of domain-specific criteria based on, e.g., the estimated amount of compensation awards obtainable in litigation. Reiterating the importance of the perceived fairness of schemes' redistributive effects, we strongly favor risk-sensitive pricing arrangements. We also call for considering a number of risk management techniques available in the financial regulatory domain to gauge contributors' risk to FGSs as a possible model to overcome hurdles of complexity.  
 
\emph{Compensation process}: In terms of the process by which victims and otherwise aggrieved parties may obtain compensation, Pearl's simple, non-adversarial approach---requiring clai\-mants to file a claim with an AIGS outlining the grounds for a compensation award and waiving their right to litigate upon acceptance of the award---coupled with appropriate appeal and dispute resolution mechanisms would presumably be suitable for most AI domains. 

\section{Conclusion}\label{Conc} 

With an eye on the primary objective pursued by AI innovation---enhancing inclusive economic and social welfare across the globe---this paper has exposed weaknesses in the existing system of legal liability and put forward solutions that would facilitate a smooth transition into an AI-driven society. 

One aim was to expand the existing literature by providing a comparative legal analysis spanning both civil and criminal legal domains to make the claim that foreseeability is a central prerequisite for attributing legal liability across all jurisdictions and legal domains. We then showed that there exist certain AI systems, which do not satisfy the foreseeability requirement, making it impossible to solve liability issues via conventional legal liability regimes under certain circumstances, and generating considerable legal uncertainty. We also raised economic and international relations arguments to highlight the economic and political costs of treating liability problems as a solely legal matter and of failure to resolve this problem of uncertainty in a timely manner. To assist current policy efforts to settle on a widely-accepted AI definition, we engaged in a discussion of the vices and virtues of such an approach, advocating functional, technology-specific definitions for regulatory and policy purposes with only an auxiliary role for more generic definitions. 

The recommended amendment to the legal liability system would better account for AI's social utility. The system of AIGSs---a solution outside of the purview of legal liability---would constitute a predictable and transparent framework for swift compensation of damages where litigation is not promising either because the category of harm caused by an AI system is unforeseeable and hence not imputable under current legal liability rules or because the process would be overly complex due to other legal intricacies. Prospective defendants would no longer need to fear arbitrary court decisions that stretch the limits of legal liability in dogmatically inconsistent, unpredictable ways to correct an otherwise uncompensated injustice. Potential victims and aggrieved parties would have peace of mind using AI, knowing that bringing complex and expensive actions of dubious outcome are no longer the only option to obtain compensation should something go wrong. By fair loss-spreading and clear allocation of risks among potential defendants and plaintiffs or prosecutors in future AI liability suits, the proposed system of AIGSs would also support emerging markets in AI technologies, in particular foster innovation and AI's social acceptance. Moreover, if desired, the AIGSs could assume a broader role within the overall regulatory framework structuring our economies. Of course, the current virtually non-existent AI governance landscape leaves quite a few blanks in regard to AIGSs' design, but reassuringly, experience with FGSs could inform many design and implementation decisions AIGS designers are likely to face. In sum, drawing on best-practice mechanisms in financial regulation, AIGSs would provide legal certainty in dealing with AI-related liability issues without violating existing liability doctrines and induce a legal environment that fosters safe and responsible AI innovation and adoption in society. 

In line with our objective to point out and raise awareness towards a general, conceptual legal problem and also because the scope of the present paper did not allow for addressing the relevant case law of multiple jurisdictions, we expressly leave this work for future research.

\bibliography{LiaBib.bib}
\bibliographystyle{plain}

\end{document}